\documentclass[fleqn,usenatbib]{mnras}

\usepackage{newtxtext,newtxmath}

\usepackage[T1]{fontenc}
\usepackage{ae,aecompl}

\usepackage{graphicx}	
\usepackage{amsmath}	
\usepackage{amssymb}	

\title[A weak H$_2$O signature in HD 179949 b]{A weak spectral signature of water vapour in the atmosphere of HD 179949 b at high spectral resolution in the \textit{L}-band}

\author[R. K. Webb et al.]{
Rebecca K. Webb$^{1,3}$\thanks{E-mail: r.k.webb@warwick.ac.uk},
Matteo Brogi$^{1,2,3}$, 
Siddharth Gandhi$^{1,3}$,
Michael R. Line$^{4}$,
\newauthor
Jayne L. Birkby$^{5}$,
Katy L. Chubb$^{6}$,
Ignas A. G. Snellen$^{7}$,
Sergey N. Yurchenko$^{8}$
\\
$^{1}$Department of Physics, University of Warwick, Gibbet Hill Road, Coventry, CV4 7AL, UK\\
$^{2}$INAF-Osservatorio Astrofisico di Torino, Via Osservatorio 20, I-10025, Pino Torinese, Italy\\
$^{3}$Centre for Exoplanets and Habitability, University of Warwick, Gibbet Hill Road, Coventry, CV4 7AL, UK\\
$^{4}$School of Earth and Space Exploration, Arizona State University, Tempe, AZ 85287, USA\\
$^{5}$Anton Pannekoek Institute for Astronomy, University of Amsterdam, Science Park 904, 1098XH Amsterdam, The Netherlands\\
$^{6}$SRON Netherlands Institute for Space Research, Sorbonnelaan 2, 3584 CA, Utrecht, The Netherlands\\
$^{7}$Leiden Observatory, Leiden University, Postbus 9513, 2300 RA Leiden, The Netherlands\\
$^{8}$Department of Physics and Astronomy, University College London, London, WC1E 6BT, UK
}

\date{Accepted XXX. Received YYY; in original form ZZZ}

\pubyear{2019}

\begin{document}
\label{firstpage}
\pagerange{\pageref{firstpage}--\pageref{lastpage}}
\maketitle

\begin{abstract}
High resolution spectroscopy ($R\,\geqslant\,20,000$) is currently the only known method to constrain the orbital solution and atmospheric properties of non-transiting hot Jupiters.
It does so by resolving the spectral features of the planet into a forest of spectral lines and directly observing its Doppler shift while orbiting the host star. In this study, we analyse VLT/CRIRES ($R=100,000$) \textit{L}-band observations of the non-transiting giant planet HD 179949 b centred around 3.5\,$\micron$. 
We observe a weak (3.0\,$\sigma$, or S/N\,=\,4.8) spectral signature of H$_{2}$O in absorption contained within the radial velocity of the planet at superior-conjunction, with a mild dependence on the choice of line list used for  the modelling. Combining this data with previous observations in the \textit{K}-band, we measure a detection significance of 8.4\,$\sigma$ for an atmosphere that is most consistent with a shallow lapse-rate, solar C/O ratio, and with CO and H$_{2}$O being the only major sources of opacity in this wavelength range. As the two sets of data were taken three years apart, this points to the absence of strong radial-velocity anomalies due, e.g., to variability in atmospheric circulation. 
We measure a projected orbital velocity for the planet of $K_\mathrm{P} = (145.2\pm 2.0)$ km\,s$^{-1}$ (1\,$\sigma$) and improve the error bars on this parameter by $\sim$70\%. However, we only marginally tighten constraints on orbital inclination ($66.2^{+3.7}_{-3.1}$ degrees) and planet mass ($0.963^{+0.036}_{-0.031}$ Jupiter masses), due to the dominant uncertainties of stellar mass and semi-major axis. 
Follow ups of radial-velocity planets are thus crucial to fully enable their accurate characterisation via high resolution spectroscopy.
\end{abstract}





\section{Introduction}

The vast majority of atmospheric characterisations of exoplanets thus far have been for transiting systems of short-period hot Jupiters using photometry and low resolution spectra \citep[e.g.][]{Sing2016}. Hot Jupiters are intrinsically more accessible for characterisation due to their extreme temperatures, $T_{\mathrm{P}}$\,>\,1000\,K, giving a relatively large ($\sim10^{-4}$) flux contrast between the planet and the parent star and larger size blocking out more of the stellar light. The molecular signatures of these hot atmospheres can be observed as extra absorption features in the transit light curve \citep{Charbonneau2002} centred on specific wavelengths for different opacity sources. Further to this, it is known that this strong irradiation on the day-side will penetrate into the deep layers of the atmosphere producing observable emitted spectra in the near-infrared \citep[NIR,][]{Seager1998}. With the continuing improvement of spectrographs, atmospheric models and analytical techniques, exoplanetary atmosphere characterisation is now at the forefront of exoplanet research.      

This past decade has seen the growth of ground based, high resolution spectroscopy (HRS) in the NIR in detecting the thermal emission from planet atmospheres \citep[for a recent comprehensive review, see][]{Birkby2018}. Such observations have provided constraints on the chemical abundances and the physical structure of the atmosphere, the first of which coming from the detection of CO in the transiting hot Jupiter HD 209458 b by \citet{snellen2010}. The success of this technique results from isolating the hundreds of individually resolved molecular lines which shift by tens of km\,s$^{-1}$ due to the large planetary velocity change over the orbit compared to quasi-stationary telluric and stellar absorption lines. There are now many methods to remove these dominating sources in the spectra, for example, through de-trending with geometric airmass \citep{Brogi2013, Brogi2014, Brogi2016, Brogi2018Giano} or with blind algorithms \citep{deKok2013svd, Piskorz2016, Piskorz2017, Birkby2017}. The cross-correlation technique with model atmospheric templates has now proved to be a robust technique in order to amplify the weak planet signal hidden within the noise of the spectra. 

HRS has now lent itself to many detections of molecular species, most of which have come from absorption of the dominating opacity sources, CO \citep[e.g.][]{Brogi2012} and H$_{2}$O \citep[e.g.][]{Birkby2013}. The resulting planet signal peak in the cross-correlation function has also allowed many physical parameters of the planet to be determined, such as, high-altitude winds \citep{snellen2010, Wyttenbach2015, Louden2015, Flowers2019}, spin rotations \citep{snellen2014, Brogi2016, Schwarz2016} and mass loss rates \citep{Nortmann2018, Allart2018}. More recently, HRS has been used for the first time to infer the presence of a strong thermal inversion from the detection of the strong optical and UV absorber TiO \citep{high_res_TiO} in the transmission spectrum of WASP-33 b. Also, HRS transmission observations of the ultra-hot Jupiter KELT-9 b has detected several ionised and neutral metal lines in this highly irradiated atmosphere \citep{Hoeijmakers2018, Cauley2019} with possible evidence for a large out-flowing, extended atmosphere \citep{Hoeijmakers2019}. 

HRS is a particularly powerful tool when observing the thermal emission from non-transiting systems on short-period orbits. Currently, this is the only known method to directly detect the orbital motion of these planets as it passes through superior conjunction, breaking the inherent degeneracy with the orbital inclination of the system and, hence, providing an accurate determination of the absolute mass of the planet. Since the probability of having a transiting system in our local neighbourhood of main sequence stars is small, HRS could offer a means of characterising the majority of these systems, particularly for very close-by systems in the habitable zone, such as Proxima Cen b \citep{Anglada-Escude2016}. However, only a handful of hot Jupiters have thus far have been characterised in this way, primarily in the \textit{K} \citep{Brogi2012, Rodler2012, Brogi2013, Brogi2014, Guilluy2019} and \textit{L}-bands \citep{Birkby2013, Lockwood2014, Piskorz2016, Piskorz2017, Birkby2017}.

In this study, we are revisiting the non-transiting system HD 179949 from previous HRS characterisation \citep[][hereafter BR14]{Brogi2014} by observing the day-side of the planet at longer wavelengths (in the \textsl{L}-band centred around 3.5\,$\micron$) with the intention of potentially observing further C and O-bearing species. This is the first time a search for molecules at 3.5\,$\micron$ is reported from HRS observations, and it tests the prediction made by \citet{deKok2014} that further species should have stronger cross correlation signals than at 2.3\,$\micron$, in particular H$_{2}$O, CH$_{4}$ and CO$_{2}$. The detection of these species and measurement of their abundances can constrain the C/O ratio in the planetary atmospheres \citep{Madhusudhan2012, Line2014, Brogi2014}, which can in turn provide insights on the formation \citep{Madhusudhan2011b} and evolution of the planetesimal in the protoplanetary disk \citep{Oberg2011}. The C/O ratio has also been used to predict whether thermal inversions are likely to be present in hot Jupiters \citep{Madhusudhan2011a, Madhusudhan2011b}. Before outlining the rest of the paper, we will give an overview of the HD 179949 system. 

\subsection{Previous observations of the HD 179949 system} \label{sec: prev observations}

HD 179949 is an F8 V \citep{Gray2006} spectral type star on the main sequence. It is slightly larger than the Sun with a mass and radius of ($1.181^{+0.039}_{-0.026})$\,M$_{\odot}$ and ($1.22^{+0.05}_{-0.04})$\,R$_{\odot}$ \citep{Takeda2007} and roughly half its age. The system is in relatively close proximity to the solar system at ($27.478 \pm 0.057$)\,pc \citep{GaiaDR2} and is bright in the NIR with a magnitude of 4.936\,$\pm$\,0.018 in the \textit{K}-band \citep{Cutri2003}. Also, due to the relatively high effective temperature of the star \citep[$T_\mathrm{eff}\approx\,6260$\,K,][]{Wittenmyer2007}, there are very few strong absorption lines observed \citep{Carpenter2009} in the infrared stellar spectrum making it an ideal target for thermal emission HRS observations.

HD 179949 b was first discovered from a radial velocity survey \citep{Tinney2001} of bright, near-by stars, with follow up photometric surveys finding no evidence of a transit. The planet was determined to have a periodicity of $P\,=\,(3.092514\,\pm\,0.000032)$ days with a semi-major axis of $\mathrm{a}\,=\,(0.0443\,\pm\,0.0026)$\,au. Due to the initial uncertainty of the inclination of the system, only a minimum mass of $M_{\mathrm{P}}\sin{i}$\,=\,$(0.916 \pm 0.076)$\,M$_{\mathrm{J}}$ \citep{Butler2006} could be determined. Subsequent analysis of mid-IR phase variations using the IRAC instrument on \textit{Spitzer} by \citet{Cowan2007}, indicated that the planet recirculates less than 21 per cent of the incident radiation to the night-side, this allows an estimate of the day-side equilibrium temperature to be $T_{\mathrm{eq}}\,\approx\,1950$\,K. Previous HRS analysis on this planet was done in the \textit{K}-band by BR14, detecting CO (S/N\,=\,5.8) and H$_{2}$O (S/N\,=\,3.9) in absorption on the day-side of the atmosphere. As such, the amplitude of the orbital velocity of the planet was found to be $K_{\mathrm{P}}$\,=\,$(142.8\,\pm\,3.4)$\,km\,s$^{-1}$, breaking the $\sin{i}$ degeneracy giving an orbital inclination of $i$\,=\,($67$\,$\pm$\,4.3)$^{\circ}$ and an absolute mass of $M_{\mathrm{P}}$\,=\,(0.98\,$\pm$\,0.04)\,M$_{\mathrm{J}}$. That analysis also found no evidence for a thermally inverted $T$-$p$ profile and a weakly constrained oxygen-rich atmosphere (C/O\,$=0.5^{+0.6}_{-0.4}$) due to a non-detection of CH$_{4}$. 

In the following sections we will give an overview of the observations in Section~\ref{sec: observations} and the data reduction in Section~\ref{sec: analysis}. We follow with the results obtained in the \textit{L}-band in Section~\ref{sec: L-band results}. We then revisit the \textsl{K}-band analysis by combining it with the \textit{K}-band data in Section~\ref{sec: combined results}. Finally, we will produce a discussion and give conclusions on this analysis in Sections~\ref{sec: discussion} and \ref{sec: conclusions}.


   
\section{Observations} \label{sec: observations}

High resolution spectra (R\,$\approx$\,$10^{5}$) of HD 179949 b were taken with the Cryogenic Infrared Echelle Spectrograph \citep[CRIRES,][]{CRIRES} on the Very Large Telescope (VLT) over two nights, 2014 April 26 and 2014 June 8. In order to achieve the highest resolving power of CRIRES, the instrument was set up using the 0.2" slit and to maximise throughput, the MACAO \citep{Arsenault2003} adaptive optics system was used.

1-D spectra were imaged on the four CRIRES CCD detectors (1024\,$\times\,$512 pixels) in the standard ABBA nodding pattern along the slit for accurate background subtraction. The spectra covered a wavelength range of 3.459-3.543\,$\micron$, giving a sampling precision of $\sim$\,1.5\,km\,s$^{-1}$\,pixel$^{-1}$. On the first night, forty spectra were taken from 2.4\,h of observation ($\phi\,=\,0.528\,-\,0.560$). The second night was split into two separate observations taken 1\,h apart, totalling 4.7\,h of observation,  with forty ($\phi$\,=\,0.397\,-\,0.428) and thirty-nine ($\phi$\,=\,0.440\,-\,0.471) spectra taken, respectively. This gives a total of 119 spectra split into three sets of $4\,\times\,\mathrm{n}_{\mathrm{frames}}\,\times\,1024$ spectral matrices, where $\mathrm{n}_{\mathrm{frames}}$ is the number of exposures (couples of AB or BA spectra) taken. Each spectral image was extracted using the CRIRES pipeline v2.3.2 and calibrated from the calibration frames that are taken the morning after the set of observations. Master dark and flat fields were created, with the inclusion of the non-linearity coefficients on the latter, to correct for detector defects and the "odd-even" effect which is known to affect detectors one and four. Further detector effects, such as isolated bad pixels and bad regions on each detector, were viewed by eye and replaced by their spline interpolated and linear interpolated values, respectively.   

\section{Data reduction} \label{sec: analysis}
\subsection{Wavelength calibration and telluric removal}

\begin{figure*}
	\includegraphics[width=\textwidth]{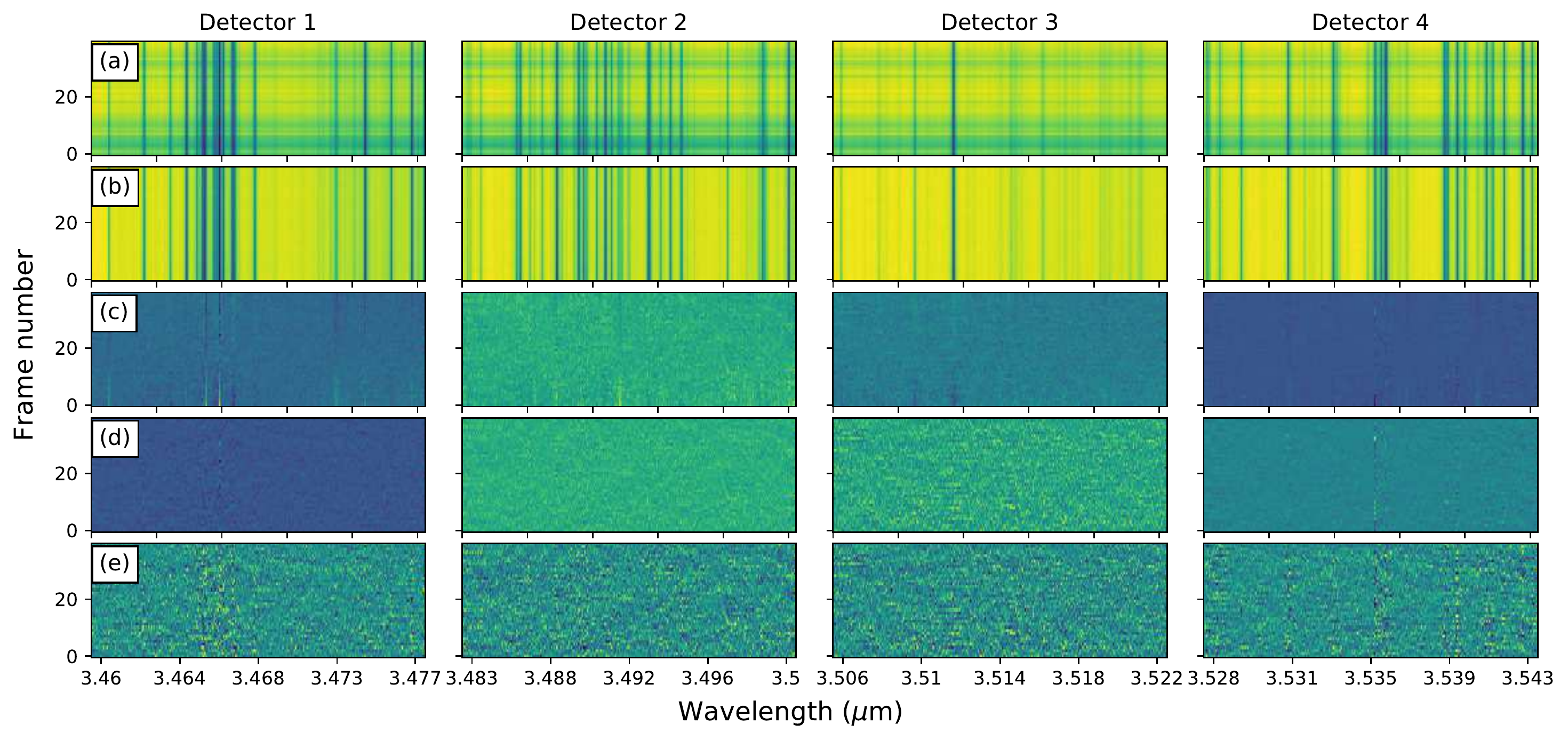}
    \caption{Example of the steps taken to remove telluric effects in the time-series of spectra taken during the first night of observations. Each column shows one of the four CRIRES detectors. Row (a): Time series of spectra extracted by the standard CRIRES pipeline, after removal of bad pixels and regions on the CCD. Row (b): Normalisation of the continuum of the spectra correcting for throughput variations. Row (c): Normalisation of the depth of the lines removing the main variability in the methane lines. Row (d): Normalisation of the time variability in the flux removing additional trends in water telluric lines. Row (e): Masking of noisy spectral channels. The same routine was applied to all of the nights observations.}
    \label{fig:telluric_steps}
\end{figure*}

In order to extract the planet's signal from the spectra, the dominating telluric contributions need to be removed.  In addition, an accurate wavelength solution needs to be determined with respect to the pixel number for each detector on each set of observations. Each stage of the analysis was performed by writing our own custom-built pipeline in \textsc{python} 3.

The most delicate part of the data reduction for CRIRES high resolution spectra has always been the alignment of the time sequence of  one-dimensional spectra to a common reference frame, and the wavelength calibration of the four detectors. In the past, this process has been done by finding the difference of the centroids of prominent telluric lines for each spectrum, shifting them through spline interpolation and comparing the spectra to a telluric spectrum with a known wavelength solution \citep{snellen2010, Brogi2012}. This approach can be costly in time and may not be practically feasible for much larger data-sets, also. Here, we fully automate this process by running a simple MCMC routine, using the \texttt{python} package \texttt{emcee} from \citet{emcee}, to determine a wavelength solution for each spectrum. This will also allow accurate error analysis on the wavelength solution. We remove detector 3 from further analysis due to the lack of prominent telluric features in these spectra which would result in an uncertain wavelength solution (see  Fig.~\ref{fig:telluric_steps}). We initialised the MCMC with three `guess' wavelengths for each spectrum which were taken to be three pixels across each detector, $x$\,=\,255, 511, 767, and their associated calibrated wavelength values from the output of the CRIRES pipeline. As in \citet{Brogi2016}, we use these three wavelengths to determine the parabolic wavelength solution of the CRIRES detectors. At each step of the MCMC, we allow the three wavelengths to randomly walk in the parameter space. Each step defines an updated wavelength solution, to which we spline-interpolate a telluric model spectrum computed via the ESO sky calculator \citep{Noll2012}. We compute the cross correlation between the telluric and the observed spectrum and convert it to a log-likelihood value using equation~(1) from \citet{Zucker2003}. This log-likelihood is used to drive the evolution of the MCMC chains.
We speed up the algorithm by running relatively short chains of a few hundreds steps multiple times and adopting their best-fit parameters as new `guess' wavelengths. Typically after the second iteration the walkers settle around the best-fit solution and this allows us to run a last, relatively short chain (12 walkers with 250 steps each in our case) which converges after a few tens of steps. The resulting wavelength solutions were found to have an average error of $0.8$\,-\,$1.8 \times 10^{-6}$\,$\micron$ which translates to an error of $0.05$\,-\,$0.1$ of a pixel and an error on the measured radial velocity of $\sim$\,150\,m\,s$^{-1}$ which was derived from the 1\,$\sigma$ quantiles of the Markov chains. Finally, we re-grid the wavelength solution to have a constant $\Delta\lambda$/$\lambda$ value and re-grid the spectra by spline interpolating to the new wavelength solution.            

Recently, it has been suggested that de-trending the data with certain methods in order to remove telluric contamination can produce spurious signals in the data \citep{Cabot2019}. As a result, we implemented two slight variations in de-trending of the data, both of which rely on removing the time dependence on the variability in the strength of the absorption lines for each spectral channel. In doing so, all the dominating  stationary absorption lines in the time-series spectra should be removed leaving the Doppler shifted planet signature largely unaltered. The first method used was to remove the linear relationship with the exponential of the airmass, directly following the de-trending method implemented by BR14. The second de-trending algorithm used here follows directly steps 3\,-\,7 from that used in \citet{Brogi2019} as shown in Fig.~\ref{fig:telluric_steps}. Panel (a) shows the data aligned on a constant $\Delta\lambda$/$\lambda$ grid, while in panel (b) we have normalised the data by dividing each spectrum by the median of the brightest 100 pixels to correct for throughput variations. In panel (c) we have divided each spectrum by a second order polynomial fit of these spectra as a function of the time averaged spectrum. While this removes most of the telluric lines, there are still residuals at the percent level, which are removed by dividing each wavelength channel through a second order polynomial fit of the measured flux as a function of time as shown in panel (d). Lastly, as in \citet{Brogi2019}, we mask noisy channels (strong telluric residuals) with a standard deviation greater than 3.5\,$\times$ of the total spectral matrix in order to use these data in a future analysis using the Bayesian atmospheric retrieval approach. We note that for future data processing through retrieval algorithms it is important to preserve the variance of each spectral channel because this enters the calculation of likelihood values directly \citep{Brogi2019}. Therefore, the common practice of `weighting' spectral channels by the variance cannot be applied, and masking is used instead. The application of two different versions of the telluric removal algorithm as outlined above was chosen to maintain consistency with BR14 while testing the performance of the more general algorithm proposed by \citet{Brogi2019}. We found that there was no significant difference for either de-trending method on the final CCFs with the  data in the following analysis and, therefore, we proceeded to only use the de-trending method used in \citet{Brogi2019}. This choice will also enable us to retrieve the atmospheric properties of the system via Bayesian analysis in the future.

\subsection{Cross-correlation analysis} \label{sec: CC}

As shown in the bottom panels of Fig.~\ref{fig:telluric_steps}, at the final stage of the analysis there remains very little residual artefacts from the spectral contaminants. However, any weak molecular signature from the planet is still hidden within the noise of the data. To observe this signal, we use a well established cross-correlation technique with several model atmospheric templates and look for any significant detection.

To match with the planet's orbital motion, the model wavelengths have to be shifted for all possible radial velocities of the planet;
\begin{equation}
    V_{\mathrm{P}} = K_{\mathrm{P}}\sin[{2\pi\phi(t)}] + V_{\mathrm{bary}}(t) + \mathrm{V_{sys}},
    \label{eq: planet vel}
\end{equation} 
accounting for the barycentric velocity of the solar system compared to Earth ($V_{\mathrm{bary}}$) as function of time $t$, and the systemic velocity of the system ($\mathrm{V_{sys}}$). In equation~\ref{eq: planet vel}, $K_{\mathrm{P}}$ is the maximum radial velocity of the planet and $\phi(t)$ are the orbital phases calculated from
\begin{equation}
\phi(t) = \frac{t - T_{0}}{P},
\label{eq: orbs}
\end{equation} 
where $T_{0}$ is the time of inferior conjunction and $P$ is the orbital period. We shifted the wavelength solution for all possible radial velocities  which was taken to be, $-249 < V_{\mathrm{r}} < 249$ km\,s$^{-1}$ in steps of 1.5\,km\,s$^{-1}$. The model fluxes were then spline interpolated, mapped onto the shifted wavelengths and cross-correlated with the observed spectra. The correlation values were then summed for all four CRIRES detectors on each night which gave three cross-correlation function (CCF) matrices in terms of time (or frame number) and radial velocity, CCF($t, V_{\mathrm{r}}$). Furthermore, we shifted these matrices to the rest frame of the planet, $V_{\mathrm{rest}}$. To do that, we needed to determine $V_{\mathrm{p}}$ from equation~(\ref{eq: planet vel}), for all orbital phases given by equation~(\ref{eq: orbs}) observed, which were computed from the orbital parameters determined in \citet{Butler2006} and from the time of observation for each spectra. In the final CCF, we weight the spectra equally as a function of phase and wavelength. Due to the uncertainty in the inclination of the system, we map out all the possible projected orbital velocities of the planet; $K_{\mathrm{P}}\,=\,0$\,-\,$200$\,km\,s$^{-1}$ in steps on 2\,km\,s$^{-1}$. The barycentric velocities were also computed from the observation times given in the fits files of each extracted spectrum. The final CCF matrix, CCF($K_{\mathrm{P}}$, $V_{\mathrm{rest}}$), was determined by co-adding the three matrices together along the time axis and dividing by the standard deviation of the total matrix, excluding values which may correspond to the planet signal, $|V_{\mathrm{rest}}|$\,<\,7.5\,km\,s$^{-1}$.

\subsection{Model atmospheres} \label{models}

The high-resolution emergent spectra models were produced from the self-consistent, line-by-line exoplanetary modelling code \textsc{genesis} \citep{Gandhi2017}. The models are produced as described in \citet{Hawker2018} and \citet{Cabot2019} resulting in a spectral resolving power of $R\,=\,300,000$ in the observed spectral band. We tested against a grid of models with the vertical atmospheric temperature-pressure ($T$-$p$) profile constructed in the same way as in BR14 for consistency. Hence, we modelled the $T$-$p$ profile by parametrising two points in space where the temperature and pressure are varied by a constant lapse rate given by,
\begin{equation}
\frac{\mathrm{d}T}{\mathrm{d}\log_{10}(p)} = \frac{T_{1} - T_{2}}{\log_{10}(p_{1}) - \log_{10}(p_{2})}.   
\end{equation}
We set the region corresponding to the planet continuum to ($T_{1}, p_{1}$) = ($1950$\,K, 1\,bar), with the upper parameters, ($T_{2}$, $p_{2}$), varied depending on the model grid used (see Tables~\ref{table: models} and ~\ref{table: multi-species models}). Above and below these regions, the atmosphere is assumed to be isothermal. We note that because the CCFs of the spectra are not weighted in this analysis (see section~\ref{sec: CC}), we approximate the day-side emission of the planet with a single $T-p$ profile and molecular abundance as an average atmospheric profile over several phases of the planet. 

We included opacity from three molecular species, H$_{2}$O, CH$_{4}$ and CO$_{2}$, into the models for the 3.5\,$\micron$ observations. The analysis by BR14 produced positive and negative detections of H$_{2}$O and CH$_{4}$, respectively, and since both species are predicted to produce more significant signals at 3.5\,$\micron$ \citep{deKok2014}, we wanted to analyse a broader range of abundances for the combined species consistent with what is expected at various atmospheric C/O ratios \citep{Madhusudhan2012}. Therefore, we generated a comprehensive grid of models (totalling 240) combining H$_2$O and CH$_4$ as described in Table~\ref{table: multi-species models}. We also included a large under-abundance, $\log_{10}(\mathrm{VMR})=-20$, for each species to simulate the absence of any opacity source from that species. We additionally also produced single molecular species models with H$_2$O and CO$_{2}$ as described in Table~\ref{table: models}. The opacity of CO$_{2}$ is expected to be lower compared to that of the CH$_{4}$ and H$_{2}$O in chemical equilibrium. However, we include CO$_2$ as the single species models allow us to analyse the data for any disequilibrium chemical processes that could produce higher abundances of observable CO$_{2}$ in the atmosphere. 

Some of the most up-to-date high resolution line list data were used for each species; CH$_{4}$ was taken from HITRAN 2016 \citep{HITRAN2017} and H$_{2}$O and CO$_{2}$ taken from the high temperature HITEMP 2010 \citep{HITEMP2010} database. We also generated single molecular models of the new and more complete water line list, POKAZATEL \citep{EXOMOL2018}, from the ExoMol database as a comparison to HITEMP regularly used in past HRS observations.

\begin{table*}
\centering
\caption{Single species grid of models analysed with the \textit{L}-band data.}
\begin{tabular}{lcccc}
\hline
Trace species & $\log_{10}$(VMR) & $T_{2}$ (K) & $\log_{10}(p_{2})$ (bars) & Line list database\\
\hline
CO$_{2}$ & [$-3.5$, $-4.5$, $-5.5$] & [1450, 1800, 2150] & [$-1.5$, $-2.5$, $-3.5$, $-4.5$] & HITEMP 2010\\
H$_{2}$O & [$-3.5$, $-4.5$, $-5.5$] & [1450, 1800, 2150] & [$-1.5$, $-2.5$, $-3.5$, $-4.5$] & EXOMOL\\
\hline
\end{tabular}
\label{table: models}
\end{table*}   

\begin{table*}
\centering
\caption{Multi-species grid of models analysed with both the \textit{L} and \textit{K}-band data. The exception with the \textit{K}-band models being that they also included a third species of CO fixed at a $\log_{10}(\mathrm{VMR})=-4.5$.}
\begin{tabular}{lccccc}
\hline
Trace species 1 & Trace species 2 & $\log_{10}$(VMR$_{1}$) & $\log_{10}$(VMR$_{2}$) & $T_{2}$ (K) & $\log_{10}(p_{2})$ (bars)\\
\hline
H$_{2}$O (HITEMP) & CH$_{4}$ (HITRAN) & [$-3.5$, $-4.5$, $-5.5$, $-20$] & [$-4.5$, $-5.5$, $-6.5$, $-7.5$, $-20$] & [1450, 1800, 2150] & [$-1.5$, $-2.5$, $-3.5$, $-4.5$]\\ 
\hline
\end{tabular}
\label{table: multi-species models}
\end{table*}  






\section{\textit{L}-band analysis} \label{sec: L-band results}

\begin{figure*}
	\includegraphics[width=\textwidth]{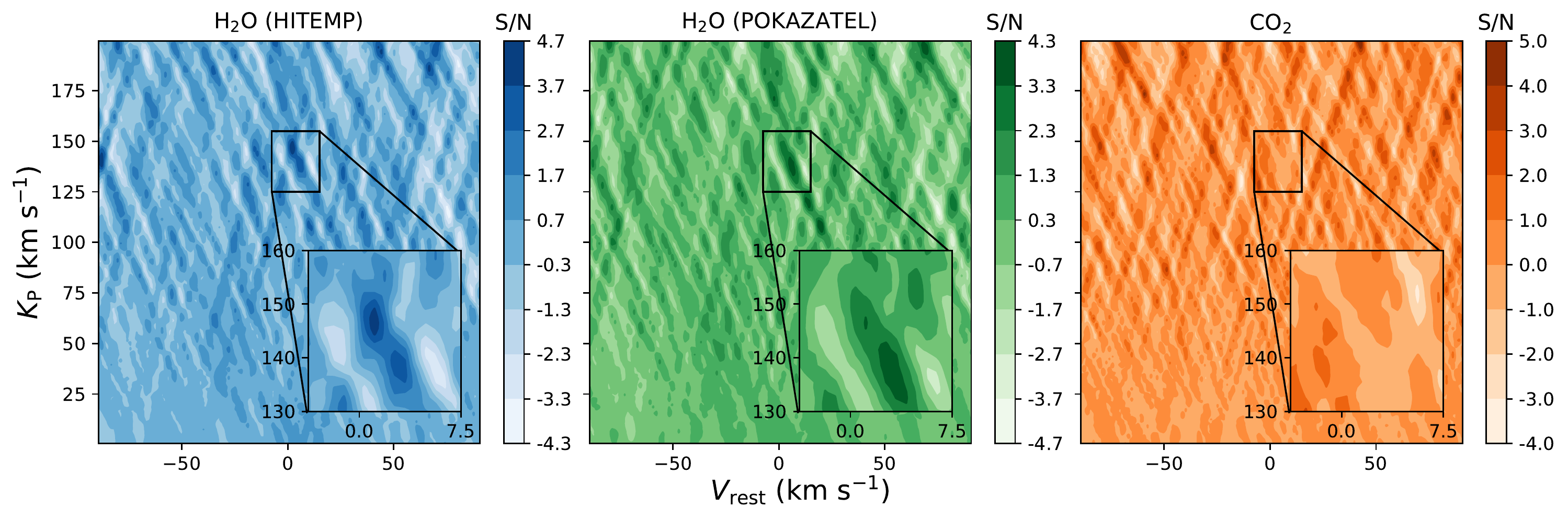}
    \caption{CCFs of all the various species analysed with the \textit{L}-band data. The velocity map is given as the projected radial velocity, $K_{\mathrm{P}}$, and the planet rest frame, $V_{\mathrm{rest}}$. The colour-bar indicates the strength in S/N of the contours. Left: The best-fitting model for H$_{2}$O and CH$_{4}$ combined models,  containing a high and negligible abundances of H$_{2}$O and CH$_{4}$, respectively, $\log_{10}(\mathrm{VMR}_{\mathrm{H_{2}O}})$\,=\,$-3.5$ and $\log_{10}(\mathrm{VMR}_{\mathrm{CH_{4}}})$\,=\,$-20$. A weak detection of H$_{2}$O can be seen in the zoomed image at $(K_{\mathrm{P}}, V_{\mathrm{rest}})$\,$\approx$\,(145, 1.5)\,km\,s$^{-1}$. Middle: CCF of H$_{2}$O with the POKAZATEL line list. There is also evidence for a weaker detection of water vapour in these models. Right: Same as the middle panel but for the models only containing CO$_{2}$. There is a non-detection for CO$_{2}$ for these models.}
    \label{fig: ccfs}
\end{figure*}

As discussed in Section~\ref{models}, we tested the \textit{L}-data against a large grid of models with various opacity sources likely to be present in the \textsl{L}\,-\,band. Each model atmosphere in the grid was cross-correlated as a function of the projected radial velocity, $K_{\mathrm{P}}$, and the systemic velocity, $V_{\mathrm{sys}}$, from equation~(\ref{eq: planet vel}). The significance of any signal in the CCF was initially taken to be the S/N, which we estimated by dividing each cross correlation value through by the standard deviation of the total CCF matrix as described in Section~\ref{sec: CC}.

In Fig.~\ref{fig: ccfs} we show the best-fitting CCFs for all the models analysed. We find evidence for a weak and localised H$_{2}$O absorption signature on the day-side emission spectrum of the planet at a maximum S/N\,=\,4.8. This signal peaks in the CCF at a $K_{\mathrm{P}}$\,$\approx$\,145\,km\,s$^{-1}$ and slightly shifted from rest frame at a $V_{\mathrm{rest}}$\,$\approx$\,1.5\,km\,s$^{-1}$. It is obtained with models with a shallow atmospheric lapse rate of $\mathrm{d}T/\mathrm{d}\log_{10}(p)$\,$\approx$\,33\,K per dex and a pure water spectrum, i.e. $\log_{10}(\mathrm{VMR}_{\mathrm{H_{2}O}})$\,=\,$-3.5$ and $\log_{10}(\mathrm{VMR}_{\mathrm{CH_{4}}})$\,=\,$-20$. It should be noted that the significance of the peak in the CCF is only weakly dependent on the $T$-$p$ profile, with a steeper profile only marginally decreasing the planet signal. Consequently, we find no evidence for CH$_{4}$ being a strong opacity source in the atmosphere, with an increasing abundance in CH$_{4}$ decreasing the strength of the planet signal from H$_{2}$O. There was also no positive correlation with the models including a inverted $T$-$p$ profile, ruling out a temperature inversion in the atmosphere HD 179949 b in agreement with BR14.   

When we analyse the data against the POKAZATEL line list grid of models in table~\ref{table: models}, we find that the CCF peak is weaker (S/N = 3.5) than the planet signal seen in the analysis with the HITEMP line list. We also find no evidence for CO$_{2}$ in the atmosphere with no significant peak in the region of the planet signal in the CCF for the entire grid of models (see the middle and right-hand plots in Fig.~\ref{fig: ccfs}).

\subsection{Expected signal retrieval with injected spectra} \label{sec: injections}

\begin{figure}
    \centering
    \includegraphics[width=\columnwidth]{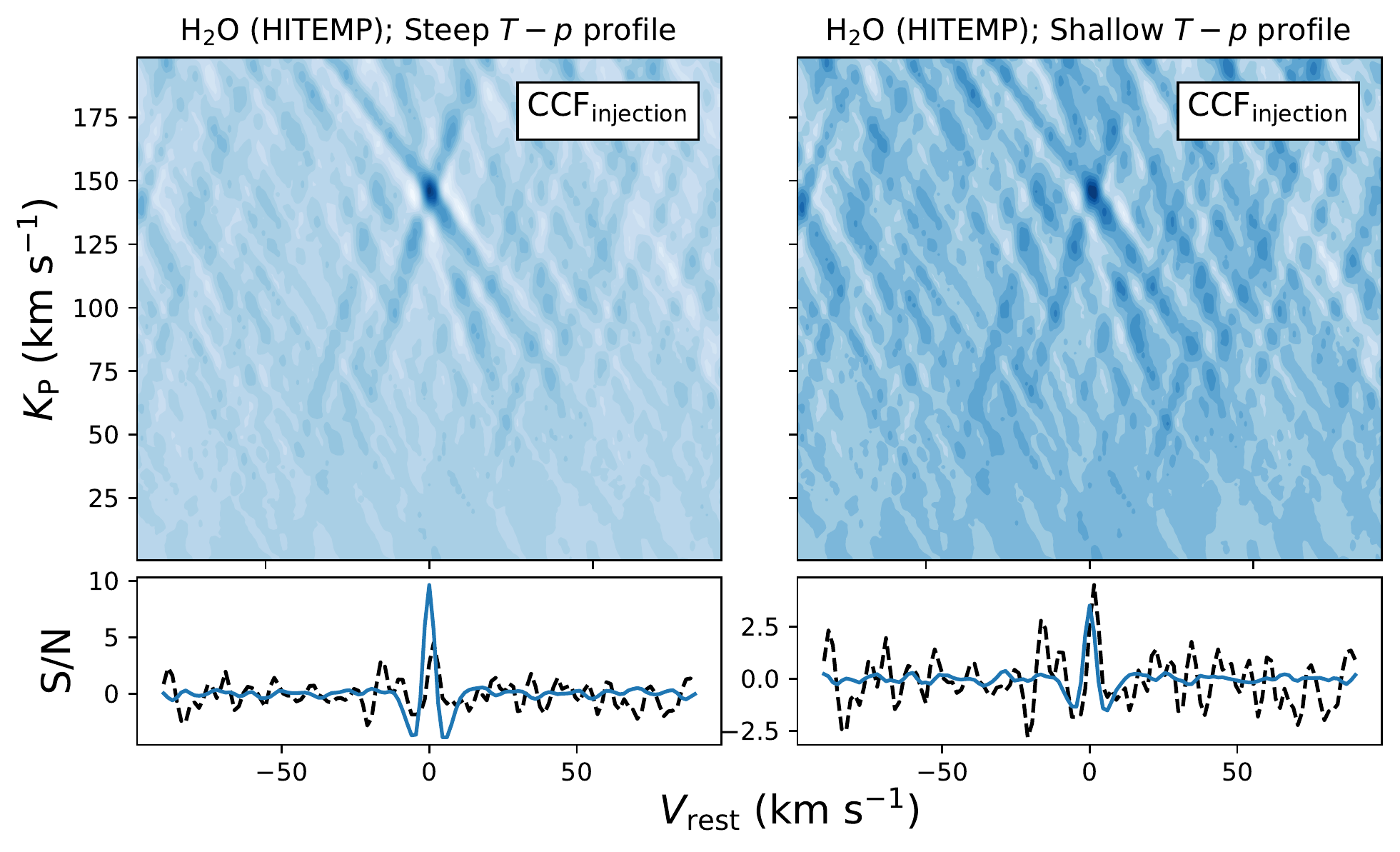}
    \caption{Injected CCFs into the \textit{L}-band data as a function of the projected radial and rest-frame velocity of the planet, $K_{\mathrm{P}}$ and $V_{\mathrm{rest}}$. Artificial spectra, pertaining to the models producing the strongest signals for the HITEMP H$_{2}$O models with no contribution from CH$_{4}$ (see Fig.~\ref{fig: ccfs}), were injected into the data (upper panels). The left and right-hand panels result from the differing steepness in $T-p$ profiles. The bottom panels show a slice of the expected (CCF$_{\mathrm{noiseless}}$, solid blue line) and observed CCFs (CCF$_{\mathrm{observed}}$, dashed black line) at the injected velocity, $K_{\mathrm{P}}$\,=\,145\,km\,s$^{-1}$. The shallower, more isothermal, $T$-$p$ profile gives us a better fit to the observed CCF.}
    \label{fig: injections(H2O+CH4)}
\end{figure}

\begin{figure}
    \centering
    \includegraphics[width=\columnwidth]{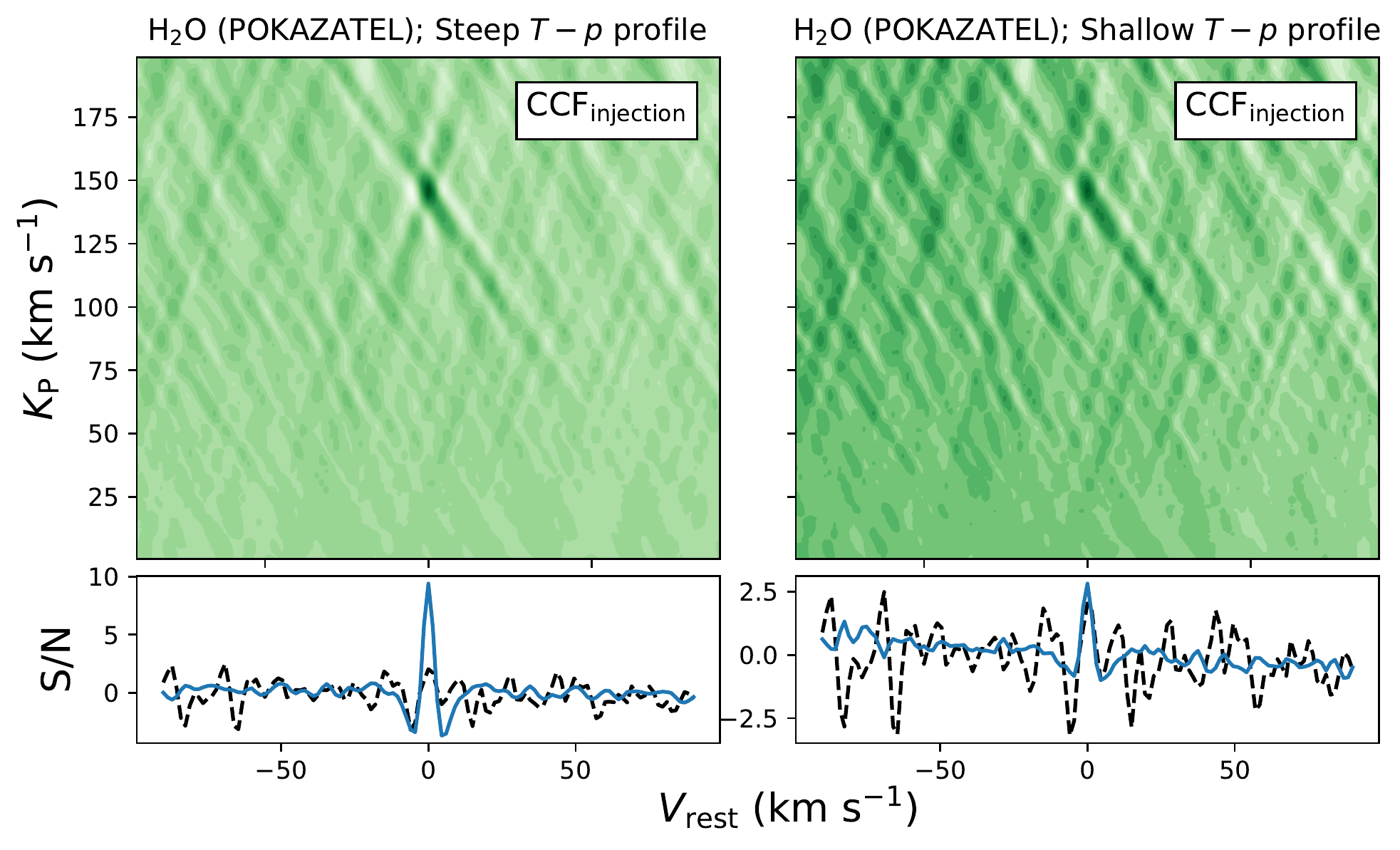}
    \caption{Same as Fig.~\ref{fig: injections(H2O+CH4)}, but for the single species POKAZATEL H$_{2}$O line list. Again, the observed CCF is more consistent with a shallower $T$-$p$ profile.}
    \label{fig: injections(POKAZATEL)}
\end{figure}

In order to give an estimation on the strength of the signal we would expect to be coming from the planet in the \textit{L}-band data, we inject artificial atmospheric spectra at the expected planet radial velocity. This gives an estimation on how sensitive this data-set is to a detection for the various species used in the atmospheric models in Tables~\ref{table: models} and \ref{table: multi-species models}. 

To extract an accurate artificial signal from the data, we first need to convert the model fluxes to the scale of observable flux values in thermal emission ($F_{\mathrm{scaled}}(\lambda)$). Here, we follow the approach from the literature \citep[e.g.][]{Brogi2014, Schwarz2015} whereby we scale each  model spectrum with the host stellar black-body ($F_{\mathrm{S}}(\lambda)$), in the wavelength range of the observations, and the ratio between the radii of the planet and star, i.e., 
\begin{equation}
    F_{\mathrm{scaled}}(\lambda) = \frac{F_{\mathrm{model}}(\lambda)}{F_{\mathrm{S}}(\lambda)} \left(\frac{R_{\mathrm{P}}}{R_{\mathrm{S}}}\right)^{2}.    
\end{equation}
The host stellar and planet parameters were taken to be; $T_{\mathrm{eff}}$\,=\,6260\,K, $R_{S}$\,=\,1.22\,$R_{\odot}$ and $R_{\mathrm{P}}$\,=\,1.35\,$\mathrm{R_{J}}$, the latter of which was also taken from the estimate given in BR14. The scaled flux was convolved to the resolution of CRIRES, spline interpolated and shifted to the planet rest frame velocity using equation~\ref{eq: planet vel}, with a fixed projected radial velocity at the position of the real planet signal observed in Fig.~\ref{fig: ccfs},  $K_{\mathrm{P}}$\,=\,145\,km\,s$^{-1}$. The artificial spectra was injected into the observed spectra ($F_{\mathrm{observed}}$) given by,
\begin{equation}
     F_{\mathrm{scaled + observed}}(\lambda) = F_{\mathrm{observed}} \times (1 + F_{\mathrm{scaled}}),
     \label{eq: superposition fluxes}
\end{equation}
as a means to include the noise structure of the observations. As a final step, these spectra are passed through the telluric removal stage of the pipeline, as described in section~\ref{sec: analysis}, before they are cross-correlated with the model spectrum that correspond to their injected spectrum.

The final CCFs for the artificially injected signals will then contain a superposition of the actual observed spectra ($\mathrm{CCF}_{observed}$) with that of the injected spectra ($\mathrm{CCF_{injection}}$) due to the inclusion of the observed spectra as indicated in equation~\ref{eq: superposition fluxes}. 
\begin{equation}
\mathrm{CCF_{noiseless}} = \mathrm{CCF_{injection}} - \mathrm{CCF_{observed}}\,,
\label{eq: CCF difference} 
\end{equation}
producing an almost noiseless CCF. We also note that because the artificial planet signal is injected into the observed spectra, we are still dividing through the cross-correlation values with the noise of the observed spectra, hence, the amplitudes of the CCFs are expressed in S/N units as in section~\ref{sec: L-band results}. 

In Fig.~\ref{fig: injections(H2O+CH4)}, we show the injected CCFs from the combined H$_{2}$O and CH$_{4}$ model that produces the strongest signal (see Section~\ref{sec: L-band results}) and compare the difference between the steep and shallow $T-p$ profiles, $\mathrm{d}T/\mathrm{d}\log_{10}(p)$\,$\approx$\,110 and 33\,K per dex, respectively. The weak planet signal seen in the CCF is more consistent with a shallower and therefore a more isothermal $T-p$ profile. The slight shift in $V_{\mathrm{rest}}$ from the observed signal  can clearly be seen when compared to the injected CCF. The width of the observed signals is qualitatively consistent with the FWHM of CRIRES indicating that there is no rotational broadening of the planet. From the CCF with a steeper profile we would have expected a much higher S/N than what has been observed in Section~\ref{sec: L-band results}. This is not surprising as a shallower temperature gradient would produce more muted absorption features in the emission spectrum. This differs from the results obtained in BR14, which find a steeper $T-p$ profile of $\mathrm{d}T/\mathrm{d}\log_{10}(p)$\,$\approx$\,330\,K per dex as their best-fitting atmospheric model. However, this result was also stated to be weakly dependant on the lapse rate. By inverting the molecular abundances in the combined models above (i.e. using a $\log_{10}(\mathrm{VMR}_{\mathrm{H_{2}O}})$\,=\,$-20$ and $\log_{10}(\mathrm{VMR}_{\mathrm{CH_{4}}})$\,=\,$-4.5$), we find very similar results as in Fig.~\ref{fig: injections(H2O+CH4)}, hence, the data is highly and weakly sensitive to strong CH$_{4}$ spectral features in steep and shallow $T$-$p$ profiles, respectively.

Similarly, in Fig.~\ref{fig: injections(POKAZATEL)} we show the injected CCFs for the H$_{2}$O POKAZATEL line list again for a shallow and steep $T$-$p$ profile and show the expected significance of a planet signal from the data. The tentative detection in the observed CCF is again consistent with the atmosphere having a shallow temperature gradient with the steeper $T$-$p$ profile clearly showing a strong signal. When the same procedure was repeated for the CO$_{2}$ models, however, even with the steep $T$-$p$ profiles the expected signal strengths were not above the threshold of detection of S/N\,$\geqslant$\,3 suggesting this data-set is not sensitive enough to observe this species.

\subsection{Constraints on the detectability of methane} \label{sec: limit on methane}

\begin{figure}
    \centering
    \includegraphics[width=\columnwidth]{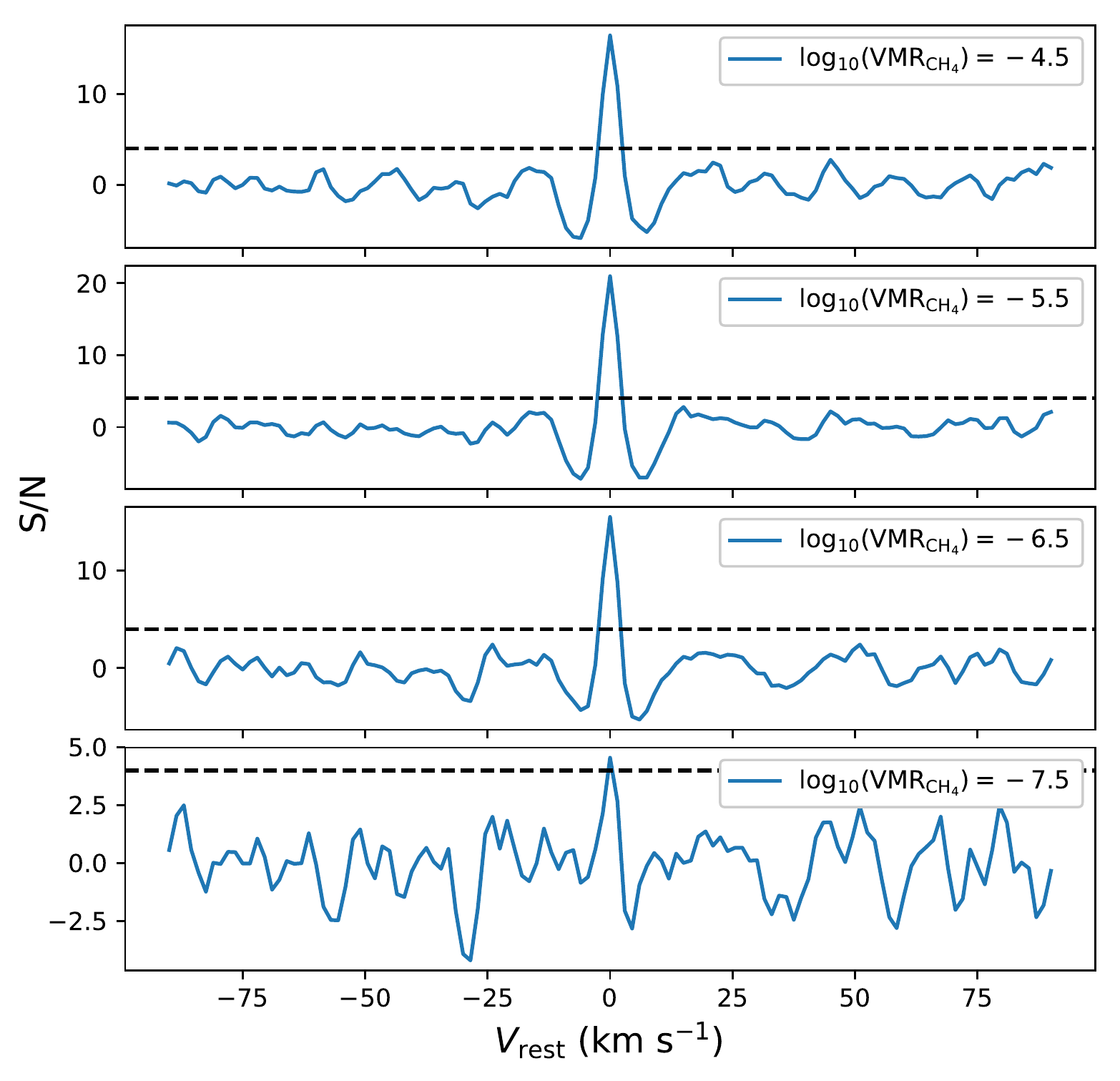}
    \caption{Injected CCFs (CCF$_{\mathrm{injection}}$) of pure CH$_{4}$ models at the atmospheric adiabatic limit, at varying abundances, into the \textit{L}-band data. The CCFs have been sliced at the injected velocity of $K_{\mathrm{P}}$\,=\,146\,km\,s$^{-1}$. The black dashed lines indicate a detection level of S/N\,=\,4.}
    \label{fig: adiabat ch4}
\end{figure}

We can also estimate the lowest abundance of CH$_{4}$ that we may be able to detect by modelling an atmosphere at the maximum possible atmospheric temperature gradient. We follow a similar analysis as in section~\ref{sec: injections} and model a spectrum of HD~179949~b at the adiabatic lapse rate for a diatomic gas, ($\mathrm{d}\ln{T}/\mathrm{d}\ln{p})|_{\mathrm{ad}}$\,=\,2\,/\,7. This lapse rate is the limit beyond which the atmosphere becomes unstable against convection. Injection and recovery of these adiabatic models with varying CH$_4$ abundances allows us to constrain the detectability.

In Fig.~\ref{fig: adiabat ch4}, we show the CCFs for the varying abundances of CH$_{4}$ sliced at the injected planet velocity. For relatively high levels of CH$_{4}$ in the atmosphere, $\log_{10}(\mathrm{VMR_{CH_{4}}})$\,$\geqslant$\,$-6.5$, we find that these signals are detectable in the CCFs peaking above the noise of the data at S/N\,$>$\,10. However, we show in the bottom panel of Fig.~\ref{fig: adiabat ch4} that for a CH$_{4}$ abundance of $\log_{10}(\mathrm{VMR_{CH_{4}}})$\,=\,$-7.5$, the CCF peaks at just above the detectable limit that we place at a S/N\,=\,4. This limit has been estimated as being  $\Delta(|\mathrm{S/N}|)$\,=\,1 above the approximate peak level of the noise of the data. At this level, we are roughly at the limit of what can be distinguished as a signal originating from the planet rather than a spurious peak in the CCF. Hence, regardless of the temperature gradient, we are unable to constrain CH$_{4}$ in the atmosphere of HD 179949 b at abundances below $\log_{10}(\mathrm{VMR_{CH_{4}}})$\,=\,$-7.5$. Chemical models of similar hot Jupiters indicate that the CH$_4$ VMR at solar abundance is $\log_{10}(\mathrm{VMR_{CH_{4}}}) \sim -7.5$ \citep{moses2013}. As the actual temperature gradient of the atmosphere of HD~179949~b is shallower than the adiabatic lapse rate, we would expect the limit of detectability to be at higher CH$_4$ abundances. Therefore, it is not unexpected that we are unable to detect CH$_4$ with these observations in the \textit{L}-band.  

\section{\textit{L} and \textit{K}-band combined analysis} \label{sec: combined results}

\begin{figure*}
    \centering
    \includegraphics[width=\textwidth]{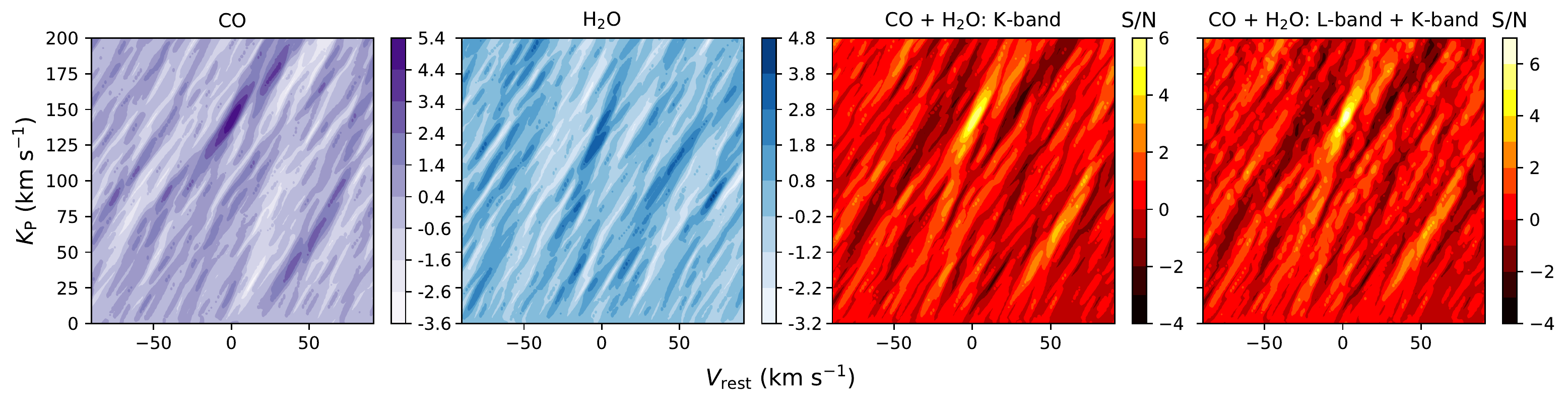}
    \caption{Best-fitting CCFs of single and combined species for the \textit{K}-band and combined data-sets. Far-left: Pure CO model CCF with the \textit{K}-band data. Centre-left: Pure H$_{2}$O model CCF with the \textit{k}-band data. Centre-right: Combined CO and H$_{2}$O species model CCF for the \textit{K}-band data. Far-right: Combined \textit{K}- and \textit{L}-band data-sets CCFs with their corresponding best-fitting combined species models (i.e. CO and H$_{2}$O and pure H$_{2}$O models for the \textit{K}- and \textit{L}-band, respectively).}
    \label{fig: combined analysis}
\end{figure*}

\begin{figure}
    \centering
    \includegraphics[width=\columnwidth]{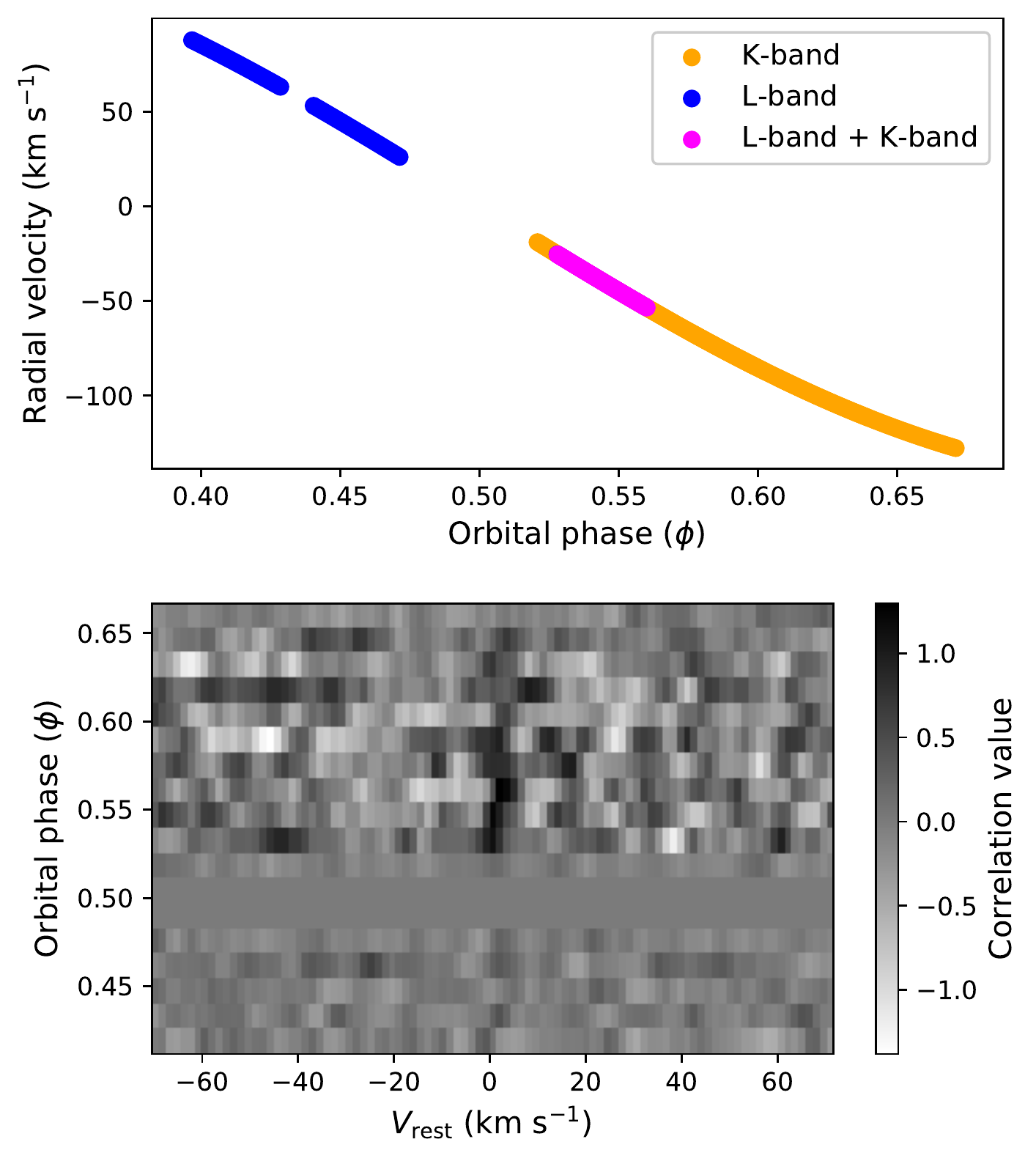}
    \caption{Top: Radial velocity of HD 179949 b as a function of the observed orbital phases in the \textit{L}-band (blue circles), \textit{K}-band (orange circles) and the phases observed with both data-sets (magenta circles). This planet radial velocity does not include the velocity corrections for an observer on earth. Bottom: Phase binned cross-correlation values of the combined data-set with both bands with their respective best-fitting model atmospheres, shifted to the planet rest-frame velocity. The gap in the right-hand panel corresponds to the large gap in the phase coverage shown in the top panel. There is a noticeable trail of positive correlation values at $V_\mathrm{rest}\approx 0$\,km\,s$^{-1}$ indicating a detection of the atmosphere of HD 179949 b.}
    \label{fig: orb ccf}
\end{figure}

We expand on the analysis by combining this data at 3.5\,$\micron$ with the previous data set observed at 2.3\,$\micron$ in order to provide better constraints on the orbital parameters of the system. We do not re-process the 2.3\,$\mu$m data here, we instead reuse the telluric-subtracted data already calculated by BR14. We also adopt their wavelength calibrations, while orbital phases are computed consistently with the previous analysis. As done in BR14, we remove detector 4 which showed residual behaviour from the known `odd-even' effect. This data-set contained a total of 500 spectra taken over three separate nights, which combined with the data taken at 3.5\,$\micron$ totals 619 spectra taken at high resolution of HD 179949 b, covering a phase range of $\phi$\,$\approx$\,(0.397-0.671) (see the left-hand panel of Fig.~\ref{fig: orb ccf}). 

To remain as consistent with the analysis done here in the \textit{L}-band and that done by BR14, we re-computed the cross correlation of the \textit{K}-band data with the models listed in Table~\ref{table: multi-species models}, and calculated with the addition of CO at a constant abundance of $\log_{10}(\mathrm{VMR_{CO}})$\,=\,-4.5. As for the \textit{L}-band data, we also estimate the S/N ratio by co-adding along the time-axis of all the spectra and dividing by the standard deviation of the total CCF matrix (see Section~\ref{sec: CC}). This was to ensure that the both data-sets were weighted equally when co-adding their correlation values. 

We are able to reproduce the results from BR14 with single species detections from CO and H$_{2}$O and a combined model of the two species as shown in the first three CCFs in Fig.~\ref{fig: combined analysis}. We also find that the best-fitting atmospheric model for HD 179949 b in the \textit{K}-band is a model containing both CO and H$_{2}$O which peaks at S/N\,=\,5.6, therefore, we include both species in the combined band analysis. We find that the best-fitting model for the \textit{K}-band data to also have a shallow lapse rate of $\mathrm{d}T/\mathrm{d}\log_{10}(p)$\,$\approx$\,33\,K per dex, with a H$_{2}$O abundance of $\log_{10}(\mathrm{VMR_{H_{2}O}})$\,=\,$-4.5$ and with no contribution from CH$_{4}$. This is fully consistent with what was found in the \textit{L}-band analysis as described in Section~\ref{sec: L-band results}. We also find that the CCFs peak at $K_{\mathrm{P}}$\,$\approx$\,143\,km\,s$^{-1}$ and at $V_{\mathrm{rest}}$\,$\approx$\,0\,km\,s$^{-1}$, as found in BR14. The final panel in Fig.~\ref{fig: combined analysis} shows the CCF of the two best-fitting models, as described in Section~\ref{sec: L-band results} and above, with the combined band data-set. This CCF peaks at a S/N\,=\,6.4 in the expected region of the planet radial velocity, $K_{\mathrm{P}}$\,$\approx$\,145\,km\,s$^{-1}$ and $V_{\mathrm{rest}}$\,$\approx$\,0\,km\,s$^{-1}$. The combination of the two bands increase the significance in S/N and further constrain the orbital signature of the planet. 

The phase resolved CCFs, binned by 0.015 in phase and spanning the orbital phase coverage for the combined data-set is shown in the bottom panel of Fig.~\ref{fig: orb ccf}. These cross correlations have been shifted to the rest frame of the planet, and positive correlation should appear as a vertical line of darker hues at $V_\mathrm{rest} \approx 0$. Indeed for certain phase bins that contain more spectra (the overlapped phase coverage seen in the top panel of Fig.~\ref{fig: orb ccf}), we see a noticeable positive correlation trail consistent with being contained within the planets radial velocity. This shows that the signal is present in both data-sets and co-adds constructively at the position of the planet, despite the difference of three years between the observations of BR14 and the $L$-band data.

\section{Statistical analysis} \label{sec: statistical analysis}

\subsection{Welch \textsl{T}-test} \label{sec: T-test}

\begin{figure}
    \centering
    \includegraphics[width=\columnwidth]{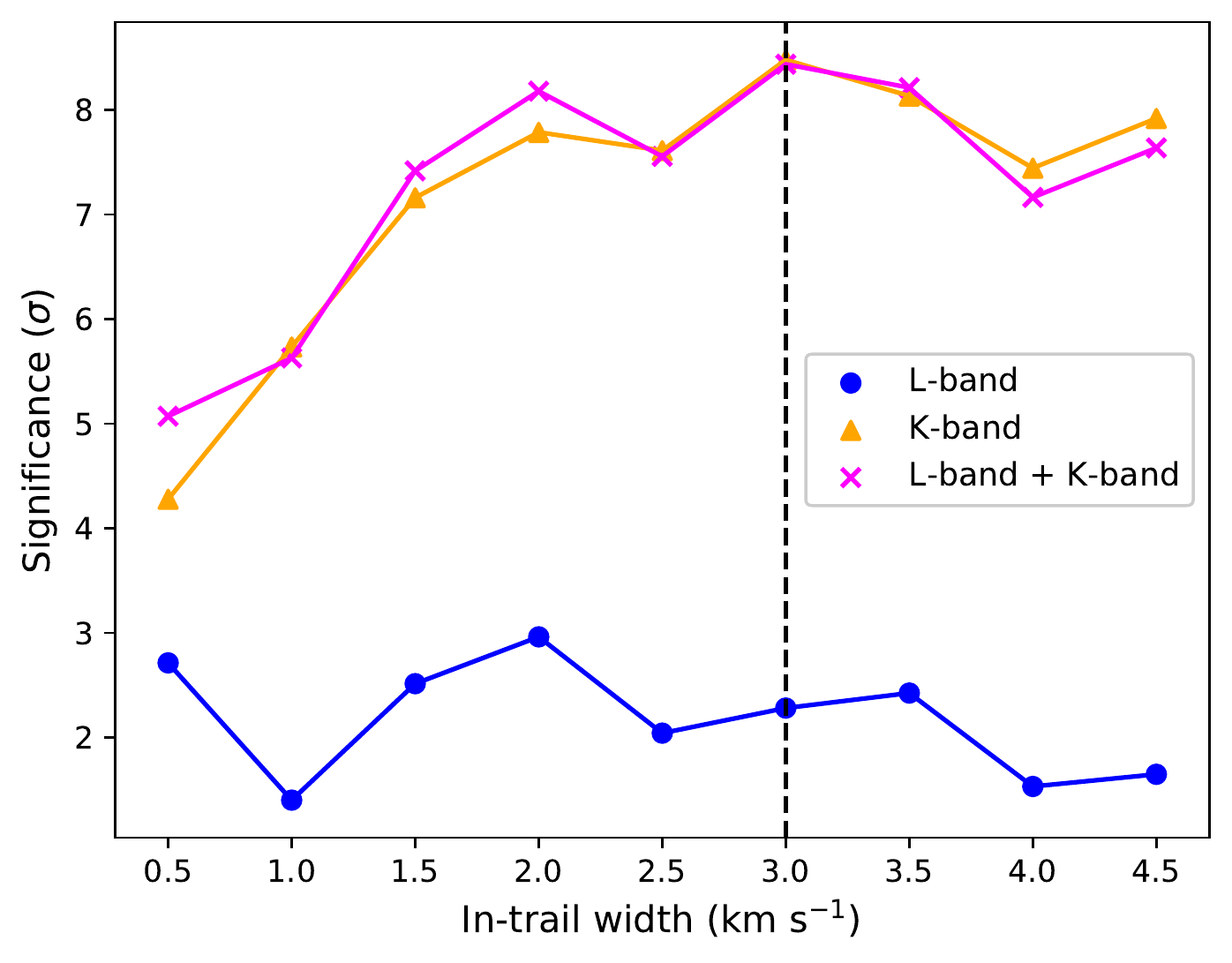}
    \caption{Welch \textsl{T}-test significance as a function of the radial velocity width included in the in-trail distributions for the best-fitting atmospheric model CCFs for each data-set. The dashed black line indicates the typical position of the FWHM of CRIRES detectors. The \textit{L}-band data peaks in significance at an in-trail width of 2\,km\,s$^{-1}$. The \textit{K}-band and combined bands peak in significance at the typical location of the FWHM for CRIRES, 3\,km\,s$^{-1}$.}
    \label{fig: sigma_rvin}
\end{figure}

\begin{figure}
    \centering
    \includegraphics[width=\columnwidth]{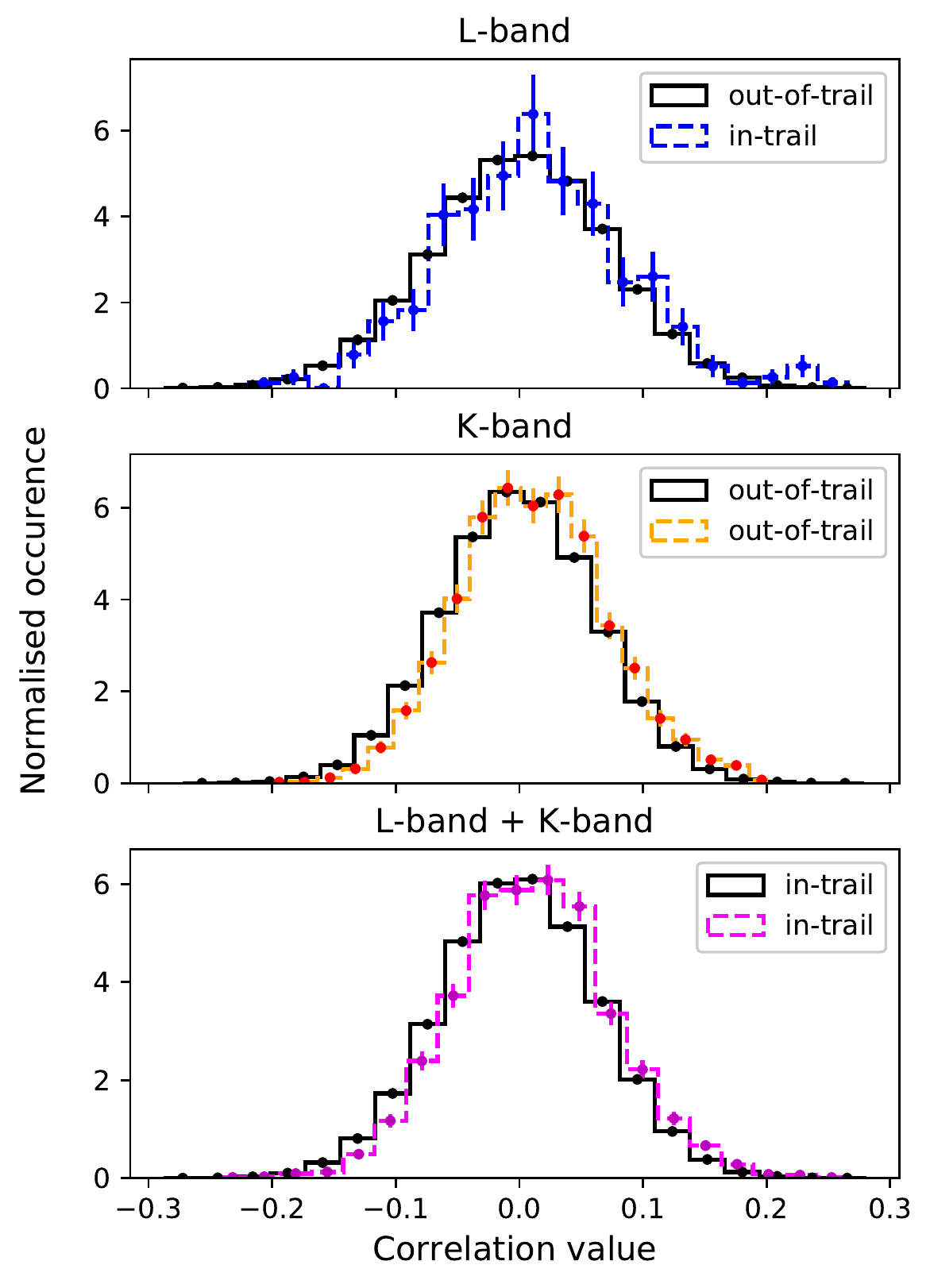}
    \caption{Normalised distributions of the correlation values within (in-trail) and outside (out-of-trail) the radial velocity of HD 179949 b for the \textit{L} (upper panel), \textit{K} (middle panel) and combined bands (lower panel). The Welch \textsl{T}-test rejects the null hypothesis for the \textit{L} (blue circles), \textit{K} (orange triangles) and combined bands (magenta crosses) by 3.0\,$\sigma$, 8.4\,$\sigma$ and 8.4\,$\sigma$, respectively. There is a noticeable positive shift in the two distributions, particularly for the \textit{K} and combined bands indicating stronger correlations for the atmospheric models within the radial velocity of the planet.}
    \label{fig: t-test}
\end{figure}

Thus far, we have only determined the significance of the CCFs by using the S/N analysis which has been shown to be a good proxy for the level of confidence for the detection of trace species in previous analyses \citep[e.g.][]{Brogi2012, Birkby2013, deKok2013svd, Brogi2016, Hoeijmakers2018, Cabot2019, Hoeijmakers2019}. However, it is usually the case in the literature to perform further statistical tests on the significance of any peaks in the CCF resulting from the signature of the planet. Apart from the standard S/N analysis, the most widely used test is the Welch \textsl{T}-test \citep{Welch1947} which is used to measure the confidence from which you can reject the null hypothesis that two Gaussian distributions that have the same mean value. We follow similar methods in the literature \citep[e.g.][]{Brogi2012} where we sample two distributions which are correlation values that fall inside and outside the radial velocity of the planet (equation~\ref{eq: planet vel}) and measure the significance that these two distributions are not drawn from the same parent distribution. We map out this significance as a function of $K_{\mathrm{P}}$ and $V_\mathrm{rest}$, as was done in the S/N analysis, and determine the $V_{\mathrm{P}}$ to be where the significance peaks in the \textsl{T}-test. We find for all bands, the detection significance peaks at the same projected radial velocity, $K_{\mathrm{P}}$\,$\approx$\,145\,km\,s$^{-1}$, therefore, we take the radial velocity to be at this value according to equation~\ref{eq: planet vel}.   

The significance of a detection that is stated by the \textsl{T}-test is strongly dependent on the chosen width of the in-trail distribution \citep{Cabot2019} and can change depending on the specific data-set and instrument used \citep{Brogi2018Giano}. We define the out-of-trail distribution to only include those correlation values more than 10\,km\,s$^{-1}$ away from the radial velocity of the planet. In Fig.~\ref{fig: sigma_rvin}, we show the dependency of the significance on the chosen radial velocity width of the planet in-trail distribution (we note that a shift of 1.5\,km\,s$^{-1}$ corresponds to $\sim$\,1 pixel on the map in Fig.~\ref{fig: orb ccf}), for each band. These are obtained from the models which give the highest S/N, i.e. a pure H$_{2}$O model ($\log_{10}(\mathrm{VMR_{H_{2}O}}) = -3.5$) and a combined model of CO and H$_{2}$O ($\log_{10}(\mathrm{VMR_{CO}}) = -4.5$ and $\log_{10}(\mathrm{VMR_{H_{2}O}}) = -4.5$) for the \textit{L} and \textit{K}-bands, respectively (see sections~\ref{sec: L-band results} and \ref{sec: combined results}). Similarly to \citet{Cabot2019}, we find that for the combined $L$- and $K$-band analysis the CCFs with the strongest signals (S/N\,$\gtrapprox$\,6) result in a much higher detection significance (8.4$\sigma$) which varies by up to 1$\sigma$ when changing the width by $\sim$\,0.5\,km\,s$^{-1}$. Vice versa, for a weak planet signal as that of the \textit{L}-band analysis, the \textsl{T}-test returns a detection significance which is 1.8$\sigma$ below the S/N level, peaking at 3.0\,$\sigma$. Overall, we obtain a peak in significance at reasonable in-trail widths of roughly the FWHM of the CRIRES detectors ($\sim$\,3\,km\,s$^{-1}$). However, the exact width of the planet signal will likely differ between data-sets because of variations in the broadening of the CCF caused by the probing of different atmospheric pressures along the optical path which is a function of wavelength. Fig.~\ref{fig: sigma_rvin} also shows that the significance of each data-set shows a steady increase to $\sim$\,1.5\,km\,s$^{-1}$, as the in-trail distributions include more of the planet signal, where the significance plateaus before decreasing again as the in-trail distribution starts to include more noise. We  note that the anomalous spike in the significance at 0.5\,km\,s$^{-1}$ ($\sim$\,3\,$\sigma$) in the \textit{L}-band data is probably due to low number statistics. Therefore, we quote to be the significance in the \textit{L}-band detection to be the peak of 3\,$\sigma$ at an in-trail velocity of 2\,km\,s$^{-1}$. 

In Fig.~\ref{fig: t-test}, we show the in- and out-of trail distributions for the two bands separately and the combined data-set. We chose the in-trail widths that peaked in significance in Fig.~\ref{fig: sigma_rvin} for each data-set. For the \textit{K} and combined bands, there is a clear sift towards higher correlation values in the in-trail compared to the out-of-trail distributions with a detection of 8.4\,$\sigma$ for both data-sets for a model containing both CO and H$_{2}$O in absorption. Qualitatively it appears that the \textit{L}-band distributions have more overlap and that is reflected in the reduced detection significance of 3.0\,$\sigma$. 

\subsection{Constraining the orbital and physical parameters of HD 179949 b} \label{sec: planet parameters}

Following the statistical testing above, we are now able to constrain the orbital and physical parameters as done in BR14. These parameters are derived from the analysis of the combined \textit{L} and \textit{K}-band data-set and their respective best-fitting atmospheric models (see Section~\ref{sec: combined results}). 

We find that the cross correlation from the best-fitting models peaks at the projected radial velocity of $K_{\mathrm{P}}$\,=\,$(145.2 \pm 2.0)$\,km\,s$^{-1}$ (1\,$\sigma$ error bars). The error bars on $K_{\mathrm{P}}$ were determined by measuring the width of the 1\,$\sigma$ contour containing the peak in the \textsl{T}-test significance map. Since we have measured directly the orbital motion of HD 179949 b with a set of time-series spectra, we can combine the orbital motion of the host star and the planet and derive the planet mass and orbital inclination of the system. As in BR14, we take the most recent measurement of the radial velocity measurement of HD 179949, $K_{\mathrm{S}}$\,=\,$(0.1126 \pm 0.0018)$\,km\,s$^{-1}$, and translate that to a mass and radial velocity ratio. Using the derived mass of HD 179949 in \citet{Takeda2007} (see Section~\ref{sec: prev observations}), this translates to an absolute planet mass of
\begin{equation}
    M_\mathrm{P} = \left(\frac{K_\mathrm{P}}{K_\mathrm{S}}\right)M_\mathrm{S} = \left( 0.963^{+0.036}_{-0.031} \right) \mathrm{M_{J}}. 
\end{equation}

Using the derived value of the semi-major axis in \citet{Wittenmyer2007}, $\mathrm{a}$\,=\,$(0.045 \pm 0.001)$\,AU, and an orbital period of $P = (3.092514\pm0.000032)$ days \citep{Butler2006}, we were able to derive the orbital inclination as:
\begin{equation}
    i = \arcsin{\left(\frac{P K_\mathrm{P}}{2\pi \mathrm{a}}\right)}=\left(66.2^{+3.7}_{-3.1}\right)^{\circ}
\end{equation} 

The error bars on both quantities were determined by drawing 10,000 random points from Gaussian distributions for the known parameters with the standard deviation equal to their quoted error bars and a mean value equal to their quoted best-fitting value. Unequal error-bars were reproduced by drawing from Gaussian distributions with unequal standard deviation for positive and negative values. Planet mass and orbital inclination were then computed as indicated above and the 15.85-84.15 per cent of the resulting empirical cumulative distribution taken as 1-$\sigma$ error bars.

Despite the revised error bars in $K_\mathrm{P}$ are 70 per cent smaller than in BR14, we were able to only slightly improve their constraints on planet mass and orbital inclination. The reason for this is that the determination of these parameters is dominated by the error on the stellar mass (for $M_\mathrm{P}$) and semi-major axis (for $i$). The parameters determined here are in full agreement within 1\,$\sigma$ with those determined in BR14. 





\section{Discussion} \label{sec: discussion}

In this study, we primarily wanted to explore the possibility that we could observe further molecular species with observations centred on 3.5\,$\micron$ from the analysis done at 2.3\,$\micron$ and, hence, improve the constraints on the C/O ratio of the planet. In \citet{deKok2014}, it is shown that at 3.5\,$\micron$, we should be able to observe H$_{2}$O, CH$_{4}$ and CO$_{2}$ with $\sim$\,2\,$\times$ the relative correlation values than at 2.3\,$\micron$, if these opacity sources are present. Furthermore, we also wanted to test the new POKAZATEL H$_{2}$O line list with the cross-correlation technique in the \textit{L}-band. Finally, we hoped to further constrain the orbital and, hence, the physical parameters of the non-transiting planet by combining the \textit{L} and \textit{K}-band data in BR14. Below, we discuss our results and the predictions made above with what we obtained in the \textit{L}-band and the subsequent merging of this data and the one presented in BR14. 

\subsection{Weak detection of water vapour in the \textit{L}-band: Astrophysical or line-list inaccuracies?}

\begin{figure}
    \centering
    \includegraphics[width=\columnwidth]{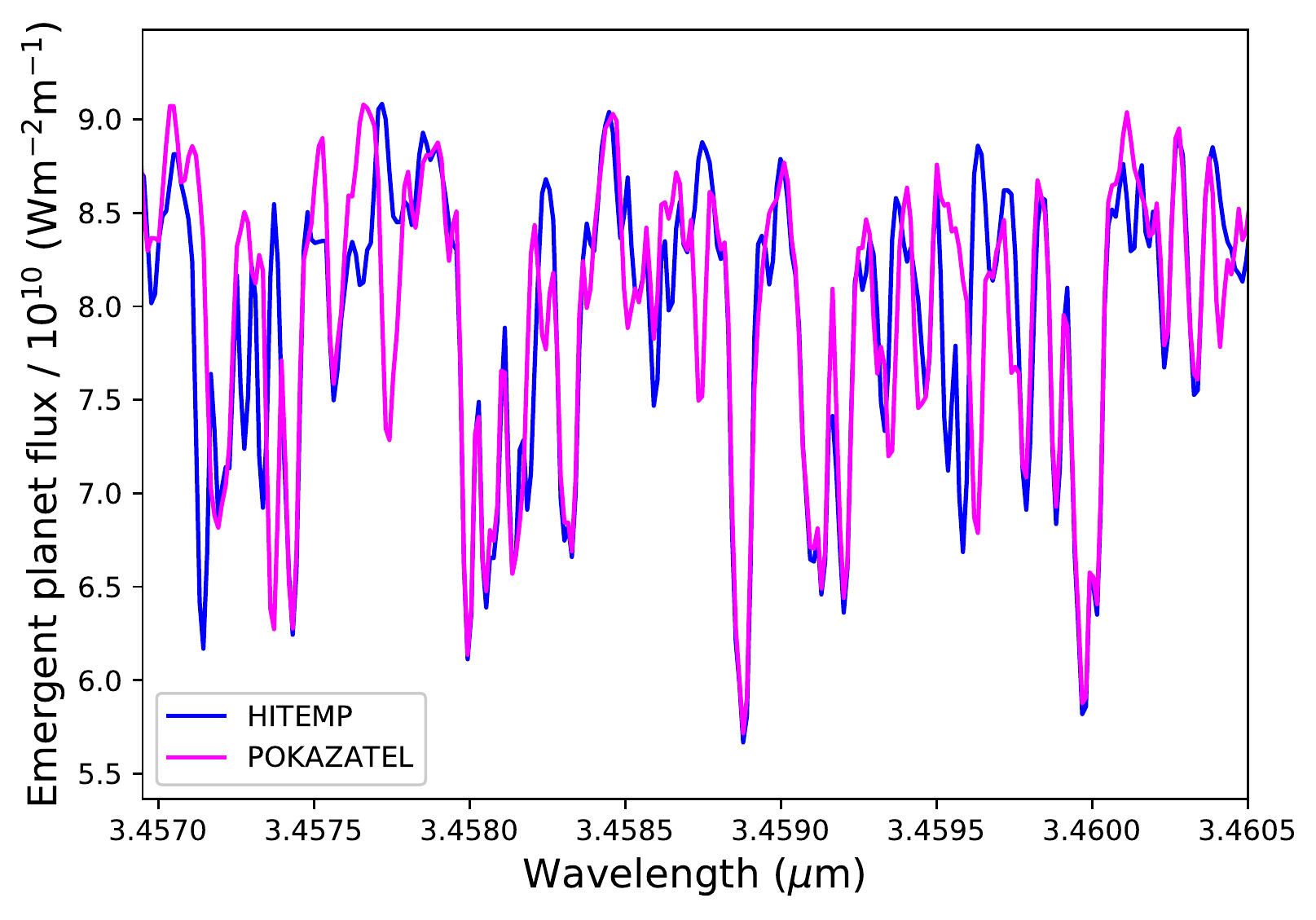}
    \caption{The model emergent planet flux in a small section of the spectral range covered in the \textit{L}-band using the HITEMP (blue) and POKAZATEL (magenta) H$_{2}$O line lists.}
    \label{fig: line list comparison}
\end{figure}

Here, we only detect a weak detection of H$_{2}$O in absorption in the thermal emission spectra of HD 179949 b at 3.5\,$\micron$ with a steep $T$-$p$ profile. We find a peak detection of H$_{2}$O in the CCF at a S/N\,=\,4.8 (see section~\ref{sec: L-band results}) which translated into a Welch \textsl{T}-test significance of 3.0\,$\sigma$ (see Section~\ref{sec: T-test}). This is perhaps on the boundary of detection significance, however, since the position of the planet signal in velocity space matches that of the strong detections in BR14, we are confident that this signal is produced by the planet and is not a spurious signal in the data. 

The question that should now be asked is why we observe in the \textit{L}-band a weaker signal than expected from BR14. In their study, it was found that the best-fitting atmospheric $T$-$p$ profile is rather steep, with a lapse rate of $\mathrm{d}T/\mathrm{d}\log_{10}(p)$\,$\approx$\,330\,K per dex. Similar lapse rates were used to drive the predictions of \citet{deKok2014}, also resulting in correspondingly stronger spectral lines. In our re-analysis of the \textit{K}-band data here, and consistently to the analysis of the \textit{L}-band, the strongest signal is found for a shallower atmospheric profile. This is further corroborated by our injection tests that seem to produce a better match to the observed amplitude of the CCF with shallow $T$-$p$ profiles. It is also predicted that highly irradiated giant planets, such as HD 179949 b, would indeed produce weaker H$_{2}$O features in the emission spectrum due to a more isothermal temperature gradient in the upper atmosphere \citep{Seager1998}. However, as mentioned in Section~\ref{sec: L-band results}, it should be noted that the cross-correlation technique is weakly dependent on the actual $T$-$p$ profile usually with only a marginal preference of the lapse rate used. By including all the models that produce a significant detection, which we chose to be within one 1\,$\sigma$ of the maximum S/N, we find a slight preference of 54 per cent for the models with the shallower lapse rate. This dual behaviour is driven by a well known degeneracy between lapse rate and abundance, with steeper lapse rates that can be accommodated by less abundant water, and vice versa. 

Previous studies have suggested that inaccuracies of line lists could hinder or even prevent detections at high spectral resolution \citep{Hoeijmakers2015}. In Fig.~\ref{fig: ccfs} we show that for the $L$-band data of HD 179949 b a signal is seen with two of the most complete line lists currently available, i.e. HITEMP and POKAZATEL, but with the latter delivering a detection weaker by a $\Delta$(S/N)\,$\sim 1$. This result is suggestive that minor differences between the line lists could play a role in this data-set too. In Fig.~\ref{fig: line list comparison}, we show a small section of the emergent planet flux in the \textit{L}-band comparing the two line lists used in this analysis at a resolution of $R = 300,000$. There are some hints that these line lists show differences at such high resolving powers in the wavelength range of these observations. This is not completely unexpected, because the cross section of water vapour around 3.5 $\micron$ is relatively weaker, and this may result in more uncertain line positioning from experimental measurements particularly for the more numerous set of weaker lines in this wavelength range. However, we do expect to extract strong signals from either line list with higher S/N observations and at wavelength bands where water is at a higher opacity than in the \textit{L}-band.

\subsection{Non-detections of carbon-bearing species}

We also analysed the \textit{L}-band data against the carbon-bearing species, CH$_{4}$ and CO$_{2}$, that, if present, would be more observable at this wavelength range. Like in BR14, we also find no evidence of CH$_{4}$ producing an observable opacity source. Injection tests with atmospheric models at the adiabatic lapse rate allow us to place a lower limit on the detectability of CH$_{4}$ at a $\log_{10}(\mathrm{VMR_{CH_{4}}})$\,=\,$-7.5$, for a minimum S/N of 4 which is our threshold for claiming a detection (see Section~\ref{sec: limit on methane}). However, even for a large abundance of CO$_{2}$, the amount of spectra obtained in the \textit{L}-band is not sensitive enough to observe this species at any physically realistic value of VMR. 

Theoretically, if we expect that the atmosphere of HD 179949 b is oxygen rich with a solar C/O ratio at chemical equilibrium (as found in BR14), then we would expect the abundances of these carbon-bearing species to be several orders of magnitude lower than H$_{2}$O \citep[e.g.][]{Madhusudhan2012, Drummond2019}. Hence, we would expect any spectral features from these additional species to be washed out by the strong opacity source of H$_{2}$O. Furthermore, this evidence of an atmospheric solar C/O ratio provides further evidence that the atmosphere does indeed have a shallow $T$-$p$ profile with the strong H$_{2}$O opacity potentially causing a strong greenhouse effect \citep{Molliere2015} in the upper layers of the atmosphere. Therefore, we attribute the non-detection of CH$_{4}$ to be likely due to the atmosphere of HD 179949 b having a solar C/O composition in chemical equilibrium. As a result we qualitatively confirm the constraints of C/O\,<\,1 provided by BR14.

\subsection{Improving the orbital parameters of the non-transiting planet HD 179949 b}

With the inclusion of the \textit{K}-band data in this analysis, we were able to improve upon the significance in S/N of the molecular signature of the planet. More importantly, we were able to improve the constraint on the projected radial velocity of the planet, $K_{\mathrm{P}}$, due to the combined observations being taken prior to and post superior conjunction. This acts to remove some of the smearing of the planet signal in the direction of whether the spectral lines are being blue or red-shifted, hence, further localising the signal in the CCF velocity map. This in turn allowed a determination on the mass and inclination of the system, however, due to the relatively large uncertainty in the stellar mass and semi-major axis, we were unable to constrain significantly better the mass of the giant planet, and we only provides a slight improvement on the inclination of the system. In line with this, all high resolution analyses on non-transiting systems thus far have also only been able to constrain the mass to the same level of uncertainty of the host stellar mass ($\geqslant$\,4 per cent) \citep{Brogi2012, Lockwood2014, Birkby2017}. Without further accurate characterisation of the stellar hosts (e.g. via asteroseismology) or follow up stellar radial velocity observations, improving the determination of planet orbital radial velocities alone using HRS with the cross-correlation technique is unlikely to significantly improve upon the determination of the mass and the inclination of the majority of non-transiting systems beyond a few percent uncertainty. 

Remarkably, we find that the radial velocities of HD 179949 b taken three years apart (2011 for the \textit{K}-band and 2014 for the \textit{L}-band) agree well and add up coherently in the rest frame of the system. Given that atmospheric circulation patterns can produce shifts up to a few km\,s$^{-1}$ in the emission spectrum of the planet \citep{zhang2017}, this means that our observations do not support any strong variability of the circulation or vertical structure of the planet over a timescale of years. Furthermore, given that for a fixed water abundance the $K$-band spectrum emerges from deeper layers of the atmosphere (higher pressure) than the \textit{L}-band spectrum, this also points to the absence of strong wind sheer between the lower and the upper portion of the day-side atmosphere. This can be seen from the lack of variability in the phase resolved CCFs (see the bottom panel of Fig.~\ref{fig: orb ccf}) for the combined data-set for this planet.
 
\section{Conclusions} \label{sec: conclusions}

In this study we have presented a follow up analysis of the non-transiting HD 179949 system using HRS in the \textit{L}-band with the CRIRES instrument. We analysed 119 spectra taken as a time series of the day-side emission. We have also produced a combined analysis with high resolution \textit{K}-band data from the previous analysis by BR14 giving a total of 619 high resolution time series spectra taken of the non-transiting planet HD 179949 b. We find a weak detection of H$_{2}$O in the \textit{L}-band with a S/N\,=\,4.8 with a Welch \textsl{T}-test significance of 3.0\,$\sigma$, the first such detection centred around 3.5\,$\micron$. We also find no evidence for any other major opacity sources in the atmosphere with this new data-set. On combining the two data-sets together, we find an improved detection significance of 8.4\,$\sigma$ for an atmosphere with CO and H$_{2}$O as opacity sources. We state this combined detection significance as the best description of this atmosphere where shielding between the individual species is likely to occur due to the different pressure levels these species absorb in the atmosphere. However, we also independently verify that we also detect CO and H$_{2}$O individually in the \textit{K}-band data as in BR14. Our best-fitting atmospheric model corresponds to a shallow lapse rate of $\mathrm{d}T/\mathrm{d}\log_{10}(p)$\,$\approx$\,33\,K per dex. This most likely explains the muted features of H$_{2}$O in the \textit{L}-band. Therefore, we find that HD 179949 b is most likely a hot Jupiter with an atmosphere that is oxygen dominated with a solar C/O ratio in chemical equilibrium that is non-thermally inverted. We also determined slight improvements on the orbital and physical parameters of the planet; $K_\mathrm{P}$\,=\,($145.2 \pm 2.0$)\,km\,s$^{-1}$ ($1 \sigma$ error contour from the Welch \textsl{T}-test), $i$\,=\,($66.2^{+3.7}_{-3.1}$)$^{\circ}$ and $M_{\mathrm{P}}$\,=\,($0.963^{+0.036}_{-0.031}$)\,$\mathrm{M_{J}}$.

We have demonstrated in this study that multiple high resolution data-sets, taken several years apart, covering different bands can be used together to characterise exoplanet atmospheres. We have also shown that by combining these data-sets can be used to improve the orbital parameters of non-transiting systems, which are inherently difficult to constrain with radial velocity measurements alone due to the uncertainty in the inclination of the system. We also find hints that, at the high resolving power of these observations, H$_{2}$O line lists may suffer from inaccuracies in line position and strength, at least in the \textit{L}-band. This is supported by the disagreement in the strength and shape of the CCFs obtained by cross correlating our data with models generated with different line lists. Although we measure a cross correlation signal from water with both line lists utilised for the modelling, we find that the strength of the signal is still dependent on the particular choice. These differences could still be relevant when the measured signals linger at the boundary of detectability, in these cases it may be necessary to use multiple line lists in order to extract the planet signal.

The recent advancements in high resolution spectrographs have and will likely provide significant improvements in HRS characterisation of exoplanet atmospheres in the future. For example, the CARMENES instrument at the Calar Alto Observatory \cite{CARMENES}, which spans over several spectral orders optical (R\,$\sim$\,94,000) and NIR (R\,$\sim$\,80,000), has recently produced a number of robust detections of transiting systems \citep{Salz2018, Allart2018, Alonso-Floriano2019a, Alonso-Floriano2019b,  Sanchez-Lopez2019}. The NIR high resolution instrument SPIRou \citep{SPIRou}, which has an even larger simultaneous wavelength coverage with a resolving power of R\,$\sim$\,73,000, is currently in operation and should also produce detections at a S/N competitive with or superior to what was possible with CRIRES. And finally, CRIRES+ \citep{CRIRES+}, which is expected to receive its first light in early 2020, will succeed the highly successful CRIRES instrument to provide improved stability and simultaneous NIR coverage by a factor of ten from its predecessor. 

\section*{Acknowledgements}

Based on observations collected at the European Southern Observatory under ESO programmes 093.C-0676(A,B,C) and 186.C-0289(N,L). 

We would like to thank the anonymous referee for their comments that have helped to improve this manuscript. We also thank Dr. Simon Albrecht and Dr. Remco de Kok for insightful discussion. MB and SG acknowledge support from the STFC research grant ST/S000631/1. MRL acknowledges support from the NASA Exoplanet Research Program award NNX17AB56G and Arizona State University Start Up funds. JLB acknowledges funding from the European Research Council (ERC) under the European Union's Horizon 2020 research and innovation programme under grant agreement No 805445. KLC acknowledges funding from the European Union's Horizon 2020 Research and Innovation Programme, under Grant Agreement 776403. IS acknowledges funding from the European Research Council (ERC) under the European Union's Horizon 2020 research and innovation programme under grant agreement No 694513.




\bibliographystyle{mnras}
\bibliography{HD179} 




\bsp	
\label{lastpage}
\end{document}